\documentstyle[11pt,newpasp,twoside,epsf]{article}

\begin{document} 

\title{Halo Substructure in
the QUEST RR Lyrae Survey} 

\author{Robert Zinn} 
\affil{Department of Astronomy, Yale University, PO Box 208101, New Haven, CT
06520-8101, USA}

\author{A. Katherina Vivas} 
\affil{Centro de Investigaciones de
Astronom{\'\i}a (CIDA), Apartado Postal 264, M\'erida 5101-A,
Venezuela}

\author{Carme Gallart}
\affil{Instituto de Astrof{\'\i}sica de
Canarias (IAC), Calle V{\'\i}a L\'actea, E-38200 La Laguna, Tenerife,
Canary Islands, Spain}

\author{Rebeccah Winnick} 
\affil{Department of Astronomy, Yale
University, PO Box 208101, New Haven, CT 06520-8101, USA}

\begin{abstract}
A survey of 380 sq. deg. of the sky with the 1m Schmidt telescope at
the Observatorio Nacional de Llano del Hato and the QUEST camera has
found 498 RR Lyrae variables lying from 4 to 60 kpc from the Sun.  We
describe the halo substructure revealed by these data and the results
of measuring some of the stars' radial velocities and metal abundances.
\end{abstract}

\section{Introduction}

Recent surveys of the galactic halo have shown that it does not have
smooth density contours but a wealth of substructure in the forms of
stellar streams and clumps.  Several of these over-densities are
unambiguously the result of the merger of one or more satellite
galaxies with the Milky Way.  While the best example of this is the Sgr dSph
galaxy and its tidal streams, evidence is accumulating for another
merger event that produced large stellar streams that wrap around the sky
(Newberg et al. 2002; Yanny et al. 2003; Ibata et al. 2003; Majewski
et al. 2003).

These discoveries confirm that the halo of the Milky Way
was built up over time by the accretion of smaller systems, as has
been suspected for decades from the range in age among halo globular
clusters and stars and from the lack of a halo metallicity gradient
(see review by Freeman \& Bland-Hawthorn 2002).  While it is no
surprise that mergers happened, the huge tidal streams that have been
revealed by the surveys are nonetheless remarkable.  Once more of the
sky has been surveyed, it may be possible to piece together the merger
history of the Milky Way and thereby document the growth of
hierarchical structure in the local universe.  The significance
that these results would have for cosmological models is discussed
elsewhere in these proceedings.

We discuss the major results of a survey for RR Lyrae variables in the
galactic halo.  Because RR Lyrae variables are easily recognized by
their characteristic light curves and periods, and because they are
excellent standard candles, they have been frequently used in the past
to map the structure of the halo.  The primary difference between our
survey and previous ones is that modern technology has permitted us to
survey a larger area of the sky and to greater limiting magnitude.
The halo substructure that we discuss here could have been discovered
long ago if for example the Lick RR Lyrae survey (Kinman et al. 1982
\& references therein) had reached $\sim 2$ mag. fainter.

\begin{figure}
\plotfiddle{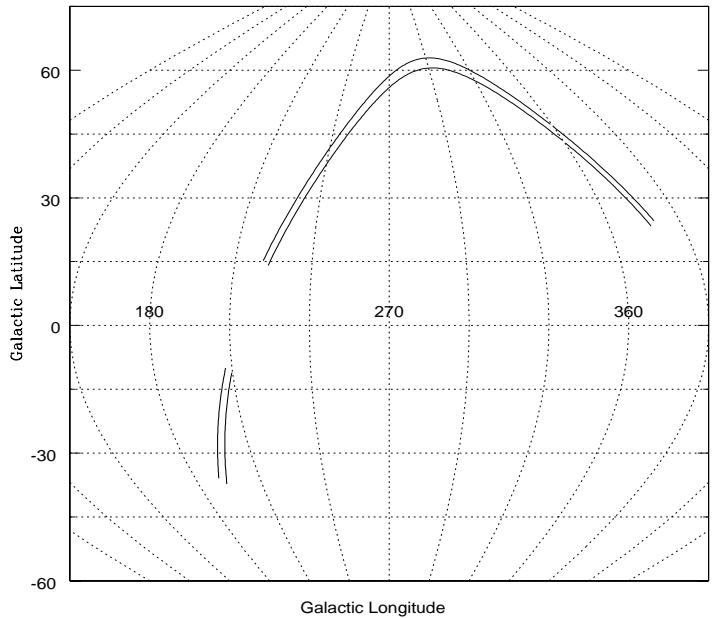}{8.0cm}{0}{50}{45}{-150}{-80}
\caption{The narrow stripe, which is $2.3^\circ$ wide, is the region
covered by the QUEST RR Lyrae survey}
\end{figure}

\section{The QUEST RR Lyrae Survey}

The QUEST survey for RR Lyrae variables was begun in 1998 as an
offshoot of a larger survey for quasars (QUEST = quasar equatorial
survey team) that was being conducted with the same instruments, the
1m Schmidt telescope at the Observatorio Nacional de Llano del Hato in
Venezuela and a large format CCD camera (Baltay et al. 2002).  The
major goals of the survey were to search a large area of the sky for
RR Lyrae variables and to use these stars to map the substructure, if
any, in the halo.  Although the Sgr dSph galaxy had been discovered
prior to the beginning of the survey, it was not clear whether the
accretion of satellite galaxies were infrequent events in the
evolution of large galaxies like the Milky Way or was the major
process that built their halos.  The survey of the first area of the
sky was completed in 2002 (Vivas 2002; Vivas et al. 2001, 2003) and is
discussed here.  The survey of a second, larger area is nearing
completion.

The observations for the survey consisted of long drift scans of the
sky that were typically several hours long in right ascension ($RA$)
and $2.3^\circ$ wide in declination ($DEC$), centered at $DEC =
-1^\circ$.  A total of 380 sq. deg. of the sky were covered by the
survey.  Figure 1 shows the galactic coordinates of the survey, which
purposely skipped the region below $10^\circ$ in latitude, because of
its large and variable interstellar extinction and its high density of
star images.

Observations in several different filters were obtained depending on
the multiple purposes of the data set, but always included the $V$
filter.  Along each line of sight of the survey, from 15 to 40 $V$
observations at intervals ranging from one day to several months were
obtained.  These observations have a saturation limit near $V=13$ and
a detection limit near $V=19.7$.  For RR Lyrae variables, these limits
correspond to distances of $\sim4$ to $\sim60$ kpc from the Sun.  The
$V$ observations were used to detect the variable stars and to determine
the periods and light curves of the RR Lyrae variables.  $V-R$ was
used to isolate candidate RR Lyrae variables by color.  The huge
amount of data gathered for this survey required that many of the data
reduction steps had to be automated, including the astrometry and the
aperture photometry.  However, the final classification of a star as a
RR Lyrae variable and the selection of its most likely period was done
by examining by eye the light curves produced by the three most
probable periods.  A few sample light curves are displayed in Vivas et
al. (2001).  Experiments with artificial variables indicate that the
survey is $>80\%$ complete for type ab RR Lyrae variables and
$40-60\%$ complete for the lower amplitude type c variables (Vivas et
al. 2003).  A total of 498 RR Lyrae variables were catalogued in the
survey region; $>86\%$ of them are new discoveries.

\begin{figure}
\plotfiddle{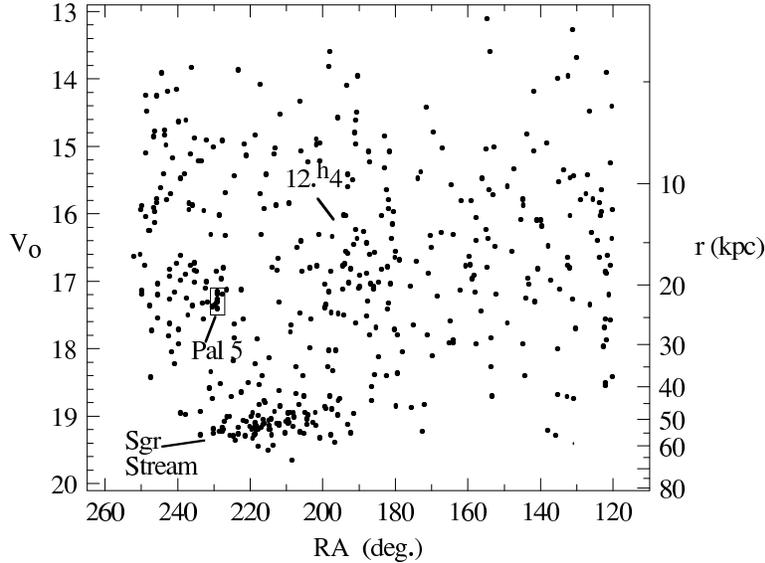}{7.0cm}{0}{80}{80}{-220}{-330}
\caption{For the northern galactic latitude region, the $V$ magnitudes
of the variables (corrected for extinction) are plotted against their
right ascensions.  On the right are the corresponding distances from the Sun.}
\end{figure}

In the following discussion, the mean apparent magnitudes of the
variables, corrected for the interstellar extinction ($V_o$), are used to
gauge their positions along the lines of sight.  It should be kept in
mind, however, that RR Lyrae variables are not perfect standard
candles and that the combination of the stellar evolution from the
zero age horizontal branch and the metallicity spread in the halo
produces a one-sigma dispersion of $\sim0.13$ in absolute $V$ magnitude
(Vivas et al. 2001).  Only on the smallest scales does this
dispersion, which is small compared to the ones for other halo
tracers, have a significant effect on the spatial distributions
inferred from $V_o$.

\section{Halo Substructure}

In Figure 2, the $V_o$ of the RR Lyrae variables in the major strip of
the survey at northern galactic latitudes (see Fig. 1) are plotted
against their $RA$ positions.  The right-hand axis gives the distances
from the Sun, assuming $M_V = +0.55$.  There are many small
clusterings in this diagram, which may be small groups of variables
that have a common origin or, more likely, a consequence of the random
fluctuations in the density distribution.  We are obtaining
spectroscopy of the densest ones to see if some of the stars have
similar radial velocities.  Three larger features, which are labelled,
have high statistical significance, and are discussed in more detail
below.

\subsection{Sgr Stream}

The large over-density of RR Lyrae variables at $V_o \sim19.2$ and
between $RA\sim 200^\circ$ and $230^\circ$ is undoubtedly part of
the tidal stream of stars from the Sgr dSph galaxy, which was
discovered in this region of the sky (Ivezic et al. 2000; Yanny et
al. 2000) prior to the completion of our survey.  Since the paper by
Vivas et al. in this volume describes this feature and our
spectroscopic observations of its stars, it will not be discussed in
detail here.

\subsection{The $12\fh4$ Clump}

Near the center of Figure 2, there is a feature that spans roughly
$25^\circ$ in $RA$ near $186^\circ$ $(12\fh4)$ and between $\sim16.5$
and 17.5 in $V_o$.  The reality of this feature is more obvious from
the plots in Figure 3 which show for three $25^\circ$ wide bands in
$RA$, which are at high galactic latitudes, the numbers of stars in bins
of $V_o$.  The most striking feature in Figure 3 is the Sgr Stream
which spans the $RA = 200-225^\circ$ zone and extends into the
$175-200^\circ$ zone (see Fig. 2).  Less prominent, but nonetheless
highly significant, is the excess of stars near the middle of the
diagram in the $175-200^\circ$ zone.  From 16.5 to 17.5 in $V_o$,
there are 35 variables in the $175-200^\circ$ zone, but only 14 and 10
in the $150-175^\circ$ and $200-225^\circ$ zones, respectively.  This
feature of halo substructure was first recognized in our data for $RA
> 195^\circ$ (Vivas et al. 2001) and comfirmed later in the Sloan
Digital Sky Survey (SDSS) for main-sequence turnoff stars (Newberg et
al. 2002), where it is referred to as the feature S297+63-20.0.

This feature may be part of the Sgr stream (Majewski et al. 2003), in
which case we expect it be similar in metallicity to the other part of
the stream we have observed (Vivas et al. this volume) and to have a
radial velocity distribution that is consistent with models of the
stream.  We are obtaining spectroscopic observations to see if indeed
this is the case.

\begin{figure}
\plotfiddle{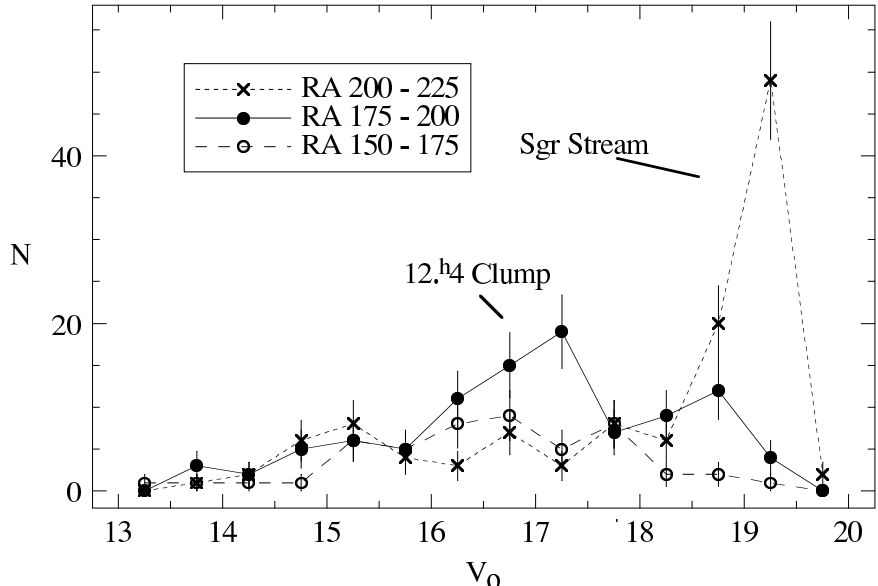}{7.0cm}{0}{90}{90}{-270}{-400}
\caption{For 3 bands of $RA$, the numbers of stars (N) in $0.4$
mag. wide bins are plotted against $V_o$.  The error bars are
$\pm\sqrt{N}$.}
\end{figure}
 
\subsection{The Pal 5 Region}

Figure 2 also reveals a large clump of stars from $RA \sim 225-250^\circ$ with
$16.6<V_o<17.6$.  Because this area includes the globular cluster Pal 5,
we refer to it as the Pal 5 region (e.g., Vivas et al. 2001).  We have
been obtaining spectroscopic observations with the 8m VLT, 3.5m WIYN,
and 1.5m ESO telescopes to measure the radial velocities and the
metallicities of these stars.  While much work still needs to be done on
this region, we report our results so far for the stars nearest
to Pal 5.

The small rectangular area labelled Pal 5 in Fig. 2, which is
$4^\circ$ wide in $RA$ and 0.4 mag. wide in $V_o$, contains the previous
known RR Lyrae variables in Pal 5, all 5 of which were recovered by
our survey.  According to our measurements, these Pal 5 variables have
a mean of 17.284 in $V_o$ and a one-sigma dispersion of 0.066.  This
dispersion is consistent with the value expected for RR Lyrae variables of
the same chemical composition lying at the same distance from the Sun.
Pal 5 is unusual in that all 5 of its variables are type c, which are
typically one-third as numerous than the type ab variables in a
cluster of the metal abundance of Pal 5.

The rectangle in Fig. 2 contains 5 additional stars (4 type ab and 1
type c), which have not been previously catalogued.  The mean and one
sigma dispersions of these stars in $V_o$ are 17.277 and 0.034,
respectively.  These values indicate that the distances to these stars
are very similar to each other and to the distance to Pal 5.  Even if one
ignores the proximity of Pal 5, these stars form such a tight
knot in space compared to the general distribution of halo RR Lyrae
variables that their physical association is highly likely.  While two
of the 5 lie relatively close to Pal 5 on the sky, the other three lie
more than one degree away, far outside the tidal radius of the
cluster.  Our spectroscopic observations of all 5 of these variables
and of 4 of the Pal 5 variables indicate that their radial velocities
and metallicities are the same as Pal 5's to within the errors.

The SDSS has detected two long stellar streams emanating from
Pal 5 (Odenkirchen et al. 2001, 2003), which is clear evidence that
this cluster is being torn apart by the tidal field of the Milky Way.
The rough outline of these streams on the sky and the positions of the
10 RR Lyrae variables in the rectangle in Figure 2 are compared in
Figure 4.  One can see that two of the 5 stars not previously
associated with Pal 5 are close to the cluster, and that one of the
three more distant stars lies in the southern stream detected by SDSS,
which is preceeding the cluster in its orbit (Odenkirchen et
al. 2003).  It is surprising that the other two distant stars do not
lie in the tidal streams, since their distances from the Sun, radial
velocities, and metallicities are consistent with membership in Pal 5.
There may be a larger system in this area of the sky, of which Pal 5 and its
streams are only part.  We are exploring this possibility by obtaining
spectroscopic observations of more RR Lyrae variables in the Pal 5
region.

\begin{figure}
\plotfiddle{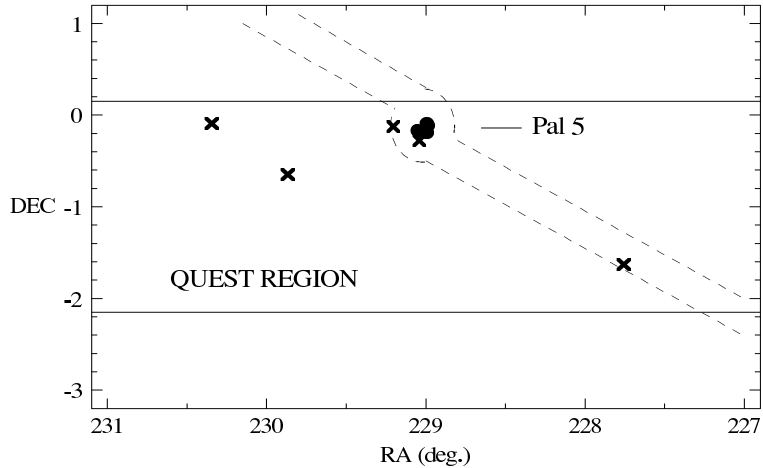}{5.5cm}{0}{70}{65}{-200}{-270}
\caption{The region near Pal 5 on the sky.  The dashed lines enclose,
approximately, the cluster and the stellar streams detected by the
SDSS (Odenkirchen et al. 2003).  The solid lines indicate the region
included in the QUEST survey.  The filled circles are the 5 previously
known RR Lyrae variables in Pal 5.  The x's depict the 5 additional
variables that appear to be related to Pal 5.}
\end{figure}

\subsection{Detection of the Monoceros Stream or Ring?}

Using SDSS data, Newberg et al. (2002) discovered a stellar feature in
Monoceros, which subsequent work has shown is part of a large
ring-like stream that appears to encircle the Milky Way (Yanny et
al. 2003; Ibata et al. 2003; Majewski et al. 2003).  The part of our
survey at southern galactic latitudes (see Fig. 1), overlaps with the
SDSS region (S200-24-19.8) where this feature was detected by Newberg
et al. (2002) as a main-sequence turnoff in the color-magnitude
diagram and where Yanny et al. (2003) confirmed its existence by
showing that the stars have very similar radial velocities.  The plot
of $V_o$ against $RA$ for the RR Lyrae variables in this region is shown
in Figure 5.  It is important to note that our survey is less complete
in this region than at northern galactic latitudes because in general
fewer observations were obtained and because the interstellar
extinction is large and variable, particularly for $RA > 80^\circ$.
While there are no large clumps of variables in Figure 5, their
distribution in $V_o$ does not appear to be random.  Ten of the 38
variables have values of $V_o$ consistent with distances from 8 to 12
kpc from the Sun, which is more than twice the number expected for a
smoothly varying distribution of $V_o$.  Yanny et al. (2003) have
estimated that in this region of the sky the stellar ring lies from
11.3 to 15.0 kpc from the Sun.  This is sufficiently close to the
distances of the excess number of RR Lyrae variables that it is likely
that they are associated.  Since Newberg et al. (2002) also detected
the ring at northern galactic latitudes from $RA = 117^\circ$ to
$130^\circ$ (feature S223+24-19.4), it is expected to be also present in
our data that cover similar latitudes starting at $RA = 120^\circ$.
As Figure 2 shows, there is a group of variables at the right $V_o$
($\sim 15.5$) to be part of the ring.  We are obtaining spectroscopic
observations to see if these variables and the ones in the $RA =
60-90^\circ$ zone have radial velocities consistent with membership
in the ring.  If confirmed as members, the RR Lyrae variables will
provide a precise estimate of the ring's distance and extent along the
line of sight, and a measurement of the metallicities of its oldest
stars.

\begin{figure}
\plotfiddle{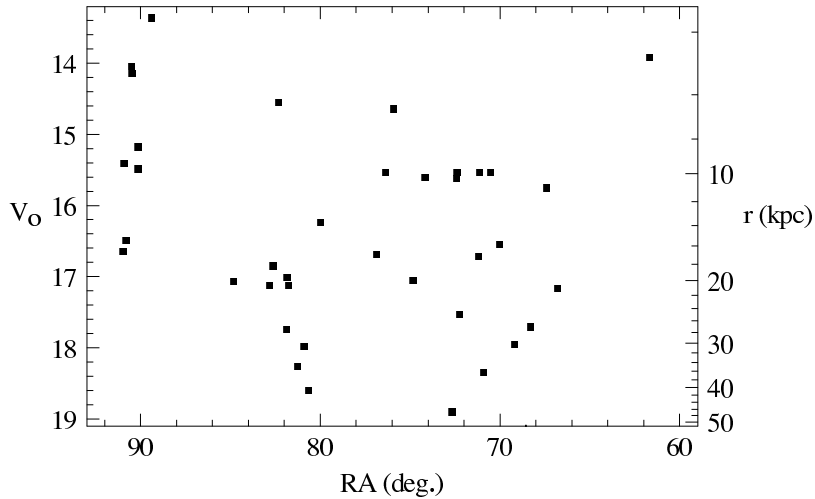}{5.5cm}{0}{80}{80}{-250}{-350}
\caption{The plot of $V_o$ against $RA$ for the survey at southern
galactic latitudes.  Distances from the Sun are shown on the
right.  The variables with $V_o \sim 15.5$ might be
associated with the stellar ring.}
\end{figure}

\section{Discussion}

The results presented above demonstrate that our RR Lyrae survey can
detect both large and small substructures.  The detection of a dense
and large-scale feature, such as the Sgr Stream, is not challenging to
a survey for any type of halo tracer that reaches sufficiently deep.
It is of course much more difficult, but nonetheless important for
documenting the merger history, to detect low density features that
may be the remains of streams from very ancient mergers or the debris
from very low-mass systems.  The detection of the excess of stars near
Pal 5 suggests that our survey has some sensitivity to low-density
features.  This is due to the combination of the high degree of
completeness of the survey and the properties of RR Lyrae variables,
which make them excellent standard candles.  Particularly in the case
of a low-density feature, it is essential to obtain spectroscopic
observations to see if the feature is also present in radial velocity
space.  We are measuring the radial velocities of many of the brighter
variables discovered by our survey to see if additional low-density
features are present.

The QUEST RR Lyrae survey is continuing, and the next band of the sky,
which is centered on $DEC = -3^\circ$, is almost complete.  We are
also beginning the Palomar-QUEST RR Lyrae survey, which employs the
Palomar 48 inch Schmidt telescope and a new CCD camera that has 112
chips.  The drift scans produced by this instrument are $4.6^\circ$
wide in DEC and have a limiting $V$ mag. of $\ge 21.0$.  For RR Lyrae
variables, this limit corresponds to a distance of $\ge120$ kpc, which
is roughly twice the distance limit of the QUEST survey.

\acknowledgments

This research was partially supported by the National Science
Foundation under grant AST-0098428.  It is part of a joint project
between Yale University and Universidad de Chile, which is partially
funded by the Fundaci{\'o}n Andes.

\end{document}